\documentclass[prl,aps,twocolumn,showpacs]{revtex4}
\usepackage{graphicx}
\usepackage{dcolumn}
\usepackage{amssymb,amsmath}
\usepackage{bm}
\usepackage{epsfig}
\def\nabbold{\mbox{\boldmath $\nabla$\unboldmath}}

\def\nbold{\mbox{\boldmath $n$\unboldmath}}

\def\beq{\begin{equation}}
\def\eeq{\end{equation}}
\def\bea{\begin{eqnarray}}
\def\eea{\end{eqnarray}}

\begin{document}
\title{Escape configuration lattice near the nematic-isotropic transition:
Tilt analogue of blue phases}
\author{Buddhapriya Chakrabarti${^1,^{a}}$}
\email{bchakrab@fas.harvard.edu}
\author{Yashodhan Hatwalne$^2$}
\email{yhat@rri.res.in}
\author{N. V. Madhusudana$^2$}
\email{nvmadhu@rri.res.in}
\affiliation{$1$ Department of
Physics,
University of Massachusetts, Amherst, MA 01003, USA. \\
$2$ Raman Research Institute, C. V.
Raman Avenue, Sadashivanagar, Bangalore 560 080 INDIA.}
\begin{abstract}
We predict the possible existence of a new phase of liquid
crystals near the nematic-isotropic ($ NI $) transition. This
phase is an achiral, tilt-analogue of the blue phase and is
composed of a lattice of {\em double-tilt}, escape-configuration
cylinders. We discuss the structure and the stability of this
phase and provide an estimate of the lattice parameter.
\end{abstract}
 \maketitle
\begin{figure}
\includegraphics[width=6cm]{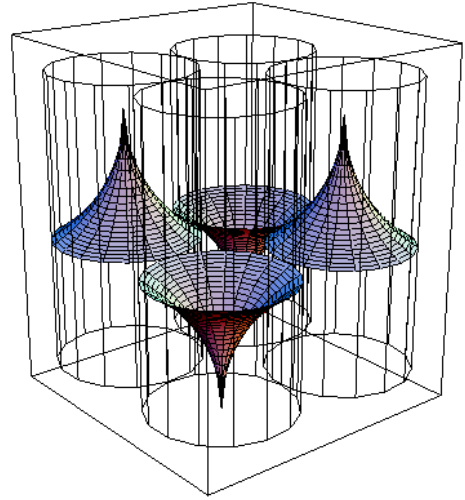}
\caption{\label{schematic} Schematic of the antiferroelectric
square lattice of double-tilt cylinders. As in the case of the
cubic blue phases, the gap between the cylinders is filled with
the higher symmetry, lower order isotropic phase. The director
field maintains continuity across the planes joining the
axes of the neighboring cylinders. See Fig.(\ref{escapecylinder})
for a two- dimensional representation of a single double-tilt
cylinder.}
\end{figure}

Liquid crystals are soft materials; this makes the existence of
phases such as the blue phases \cite{de-Gennes:93,Meiboom:81}, the
twist grain boundary phases \cite{Renn:88, Goodby:88} and the
smectic blue phases \cite{DiDonna:03,Pansu:01} possible. In this
paper we explore the possible existence of a new phase of liquid
crystals intervening the nematic ($N$) and the isotropic ($I$)
phases. The proposed phase is composed of escape-configuration
cylinders \cite{Mayer:73,Cladis:72}, and can be thought of as an
{\em achiral, tilt analogue of blue phases} (see
Fig.\ref{schematic}). Blue phases are composed of double-twist
cylinders and intervene between the cholesteric and the isotropic
phases. Interestingly, the stability of double-twist cylinders
forming blue phases is not solely due to the chirality of
molecules. Double-twist cylinders are favored over the simple
cholesteric with twist along a single axis if the coefficient of
the saddle-splay deformation term in the elastic free energy is
positive \cite{Meiboom:81,de-Gennes:93}. The saddle-splay
deformation (see (\ref{Saddle-Splay})) corresponds to a
surface-like term in the free energy (it is a total divergence and
integrates to the bounding surface by Gauss's theorem). Therefore
blue phases {\em necessarily require} the presence of interfaces
for their stability. The gaps between the double twist cylinders
in the cubic blue phases are believed to be filled with higher
symmetry, lower order isotropic material\cite{de-Gennes:93}.

\begin{figure}
\includegraphics[width=6cm]{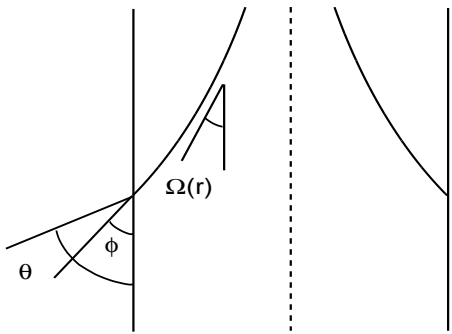}
\caption{\label{escapecylinder}Schematic of the escape
configuration. The solid vertical lines represent the
nematic-isotropic interface. The angle $\theta $ preferred by the
weak anchoring energy differs from the angle $\phi $ which
minimizes the total energy. The nematic director is tangential to
the curved lines.}
\end{figure}

For a {\em two-dimensional} nematic confined within a circle, and
subject to the boundary condition that the nematic director is
radial at the circumference, it is a topological imperative that
the director field inside the circle have point singularities
(disclinations) with a total charge of +1. However, a similar
topological requirement does not apply to a {\em
three-dimensional} nematic confined within a cylindrical capillary
with radial boundary conditions. The $ +1 $ {\em line} singularity
can be removed by the {\em escape of the nematic director into the
third dimension} \cite{Mayer:73,Cladis:72}. The singular
disclination line is stable if $ \ln(R/a) < \frac{\pi}{2}
(K_{3}/K_{1})^{1/2} $, where $ R $ is the radius of the capillary,
and $ a $ is a microscopic cut-off (see (\ref{Frank-F}) below).
Clearly, escape-configurations are favored over configurations
with topological singularities for reasonable capillary radii
unless the splay elastic constant is unusually small. The director
field $ {\hat {\bf n}} $ in the escape-configuration cylinders
(Fig.\ref{escapecylinder}) is tilted with respect to the axis of
the cylinder, and has splay as well as bend deformation. This tilt
of the director can be resolved in two mutually orthogonal
directions in a plane perpendicular to the axis of the cylinder;
they are {\em double-tilt cylinders}. These cylinders clearly have
a non-zero saddle-splay deformation: the two principal curvatures
of the nematic director field at any point in the cylinder have
opposite signs.

Can a phase composed of the double-tilt cylinders exist near the
nematic-isotropic ($NI$) transition for reasonable values of
parameters? In what follows we answer this question in the
affirmative, and propose a possible candidate for such a phase. In
the spirit of the ``low chirality'' analysis of  blue phases
\cite{Meiboom:81,de-Gennes:93}, we use the continuum description
of nematics and ignore the effects of thermal fluctuations.
As is the case with the low chirality theory, the theory presented
in this letter is not a theory of phase transitions based upon
an order parameter description.

Our  results are summarized in Figs.(\ref{schematic},
\ref{Phase-Diagram-k-vs-mu24} and \ref{R-vs-mu24}). The lattice
depicted schematically in Fig.(\ref{schematic}) is stabilized by
the competition between ($i$) the saddle-splay deformation
(\ref{Saddle-Splay}); and ($ii$) the bulk nematic deformation
(\ref{Frank-F}), the nematic-isotropic interface energy
(\ref{Surface-Tension}), and the weak-anchoring energy at the
interface (\ref{Rapini-Papoular}). In the structure proposed in
this paper, the gaps between the cylinders are filled up with the
high symmetry, low order isotropic material. Notice that the
lattice shown in Fig.(\ref{schematic}) is an {\em
antiferroelectric} lattice: neighboring cylinders have opposite
escape directions to maintain the continuity of the director field
across the planes joining the cylinder axes. Each double-tilt
cylinder can be characterized by the direction of escape (up or
down), and is a polar object even if the nematogenic molecules are
not polar. Fig.(\ref{Phase-Diagram-k-vs-mu24}) shows that the
proposed phase has a small region of stability  near the $ NI $
transition temperature in the space of the dimensionless
parameters $ \mu_{24} = K_{24}/K $, the relative strength of the
coefficient of the saddle-splay term to the typical coefficient of
the bulk Frank free energy, and $ k = \gamma/\sigma $, the
relative strength of the coefficient of the nematic-isotropic
interfacial tension to the coefficient of anchoring energy at the
interface. In the case of cubic blue phases, the pitch of the
cholesteric provides a natural length scale, and the nematic
director twists through $ \pi/4 $ from the center of the
double-twist cylinder to its boundary. This ensures that the
director is continuous across neighboring double-twist cylinders.
In the present problem, there is no ``natural" length scale at the
lowest order of the continuum description that we employ. We
therefore calculate the radius of double-tilt cylinders by solving
the free boundary condition that arises naturally from the
variational problem of minimizing the total free energy (see
(\ref{Boundary-Condition})). Fig.(\ref{R-vs-mu24}) shows the
variation of the radius of the cylinder (which is half the lattice
parameter) with $ \mu_{24} $ for various values of $ k $.

In order to derive these results we begin by listing the
contributions to the total free energy. The bulk Frank free energy
density is
\begin{eqnarray}
\label{Frank-F}
f_{\rm F} =
\frac{K_{1}}{2}(\nabbold\cdot\hat{\nbold})^{2}
& + & \frac{K_{2}}{2}(\hat{\nbold}\cdot(\nabbold\times\hat{\nbold}))^{2}
\nonumber \\
& + & \frac{K_{3}}{2}(\hat{\nbold}\times(\nabbold\times\hat{\nbold}))^{2},
\end{eqnarray}
where $ K_{1}, K_{2},$ and $K_{3}$ are respectively the splay,
twist, and bend elastic constants.  In what follows, we use the
``one-constant" approximation $ K_{1} = K_{2} = K_{3} = K $, which is
indeed a reasonable approximation near the nematic-isotropic transition
temperature \cite{footnote2}.

In addition to the bulk contribution, we have the surface-like
saddle-splay elastic free energy density
\begin{equation}
\label{Saddle-Splay} f_{\rm SS} = K_{24} \nabbold \cdot
[\hat{\nbold} (\nabbold \cdot \hat{\nbold}) - (\hat{\nbold} \cdot
\nabbold) \hat{\nbold} ].
\end{equation}
We note that the sign of $ K_{24} $ is not dictated by stability
conditions.

The free energy associated with the nematic-isotropic interface is
composed of two parts, the usual surface tension part
\begin{equation}
\label{Surface-Tension} F_{\rm ST} = \gamma \int d^2 S
\end{equation}
and the Rapini-Papoular \cite{Rapini:69} ``weak anchoring energy"
\begin{equation}
\label{Rapini-Papoular}
 F_{\rm RP} = \frac{\sigma}{2} \int \sin^2[\theta - \phi] d^2 S.
\end{equation}
where $ \theta $ is the preferred angle of the nematic director at
the interface (see Fig.\ref{escapecylinder}), and the integrals
are over the interface. Experimentally,
values of $ \theta $ as large as about $ 45^{o} $ have been
reported \cite{Faetti-Palleschi:84}. The spontaneous nonzero tilt
of the director at the interface has its origin \cite{Barbero:86} in
the order electric polarization \cite{Prost:77}
arising from the gradient of the nematic order parameter across
the nematic-isotropic interface. Since the present model is not
based upon an order parameter description, it is sufficient for our
purpose to recognize that $ \theta $ is nonzero and use
(\ref{Rapini-Papoular}) for the energy cost of deviations from the
angle $ \theta $ at the interface.
\begin{figure}
\includegraphics[width=6cm]{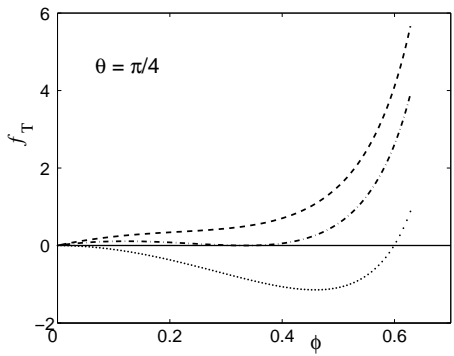}
\caption{\label{Free-Energy-vs-Phi} Variation of the free energy
per unit length of a single double-tilt cylinder as a function of
$ \phi $ for $ \theta = \pi/4 $, $ k = \gamma/\sigma = 1 $; $
\mu_{24} = - 1.3 $ (dashed line), $\mu_{24} = - 1.373 $
(dash-dotted line) and $ \mu_{24} = - 1.5 $ (dotted line). For $
\mu_{24} \simeq -1.5 $, $ f_{\rm T} $ has a minimum.}
\end{figure}

We now discuss the energetics of a single escape-configuration
cylinder {\it at} the $ NI $ transition (at the transition point, 
there is no free energy difference between the nematic in the
escape cylinders and the isotropic material filling in the gaps
between them). The total free energy for our model is
\begin{equation}
\label{Total-Free-energy} F_{\rm T} = F_{\rm F} + F_{\rm SS} +
F_{\rm ST} + F_{\rm RP},
\end{equation}
where $ F_{\rm F} = \int f_{\rm F} d^{3}x $, and
$ F_{\rm SS} = \int f_{\rm SS} d^{3}x $.

We parameterize the nematic director as 
$ n_{r} = \sin \Omega(r), n_{\phi} =  0, n_{z} = \cos \Omega(r) $ 
in cylindrical polar coordinates, where $ \Omega(r) $ is the angle 
made by the director with the axis of the cylinder 
(see Fig.{\ref{escapecylinder} ). The free energy {\em per unit length} 
of the cylinder of radius $R$ is
\begin{eqnarray}
\label{Total-Free-energy-exp} f_{\rm T} = \int^{R}_{0} [
\frac{K}{2}( (\frac{d \Omega}{d r})^2 &+& \frac{\sin^2
\Omega}{r^2}) + \frac{{\tilde K}}{r} \sin 2\,\Omega
\frac{d \Omega}{d r} ] 2 \pi r dr \nonumber \\
&+& \frac{1}{L_z} [ F_{\rm ST} + F_{\rm RP}],
\end{eqnarray}
where $ {\tilde K} = K_{24} + (K/2) $, and $L_z$ the length along
the cylinder axis is unity. The Euler-Lagrange equation reads
\begin{equation}
\label{Euler-Lagrange} \frac{d^2 \Omega}{dr^2} + \frac{1}{r}
\frac{d \Omega}{dr} - \frac{1}{r^2} \sin \Omega \cos \Omega = 0.
\end{equation}
The variational problem of minimizing the total free energy leads
to the {\em free} boundary condition
\begin{equation}
\label{Boundary-Condition} K \frac{d\Omega}{dr} \vert_{r=R} +
\frac{{\tilde K}}{r} \sin \Omega \cos \Omega \vert_{r=R} +
\frac{\sigma}{2} \sin 2(\Omega - \theta)\vert_{r=R} = 0,
\end{equation}
where $R$ is the radius of the escape cylinder. This condition
which ensures torque balance at the boundary of the escape
cylinders picks out the radius of the escape cylinders and hence
the lattice parameter.

We recognize that the  angle $ \phi $ made by the nematic director
at the $ NI $ interface is not necessarily equal to the angle $
\theta $ preferred by the anchoring energy. Therefore, we first
solve Eq.(\ref{Euler-Lagrange}) subject to the condition that the
nematic director makes an angle $ \phi $ at the $ NI $ interface,
and then proceed to minimize the total free energy with respect to
$ \phi $ (see Fig. \ref{Free-Energy-vs-Phi}).

The solution to the Euler-Lagrange equation is \cite{Doane:92}
\begin{equation}
\label{Escape-Solution} \Omega(r) = 2 \arctan \left[\frac{r}{R}
\tan \left(\frac{\phi}{2} \right) \right].
\end{equation}

The free boundary condition (\ref{Boundary-Condition}) relates
the radius $ R $ of the cylinder to the angle $ \Omega = \phi $ at the
boundary of the cylinder, so that
\begin{equation}
\label{Radius} R = \frac{\lambda}{2} \csc(\theta - \phi)
\sec(\theta - \phi)
  (2 \sin \phi + \mu \sin 2 \phi),
\end{equation}
where $ \lambda = K/\sigma $ is a length, $\mu = 2 \mu_{24} + 1 $,
with $ \mu_{24} = K_{24}/K $. Substituting the expression
(\ref{Radius}) for the radius in  $ f_{\rm T} $ yields the energy
per unit length (measured in units of $K$) of a single double-tilt
cylinder of radius $ R $:
\begin{equation}
\label{Full-Free-Energy-vs-Phi}
f_{\rm T} = 2\pi \left[ \sin^{2} \frac{\phi}{2}
+ \mu \sin^{2} \phi + (\frac{k}{2}
+ \frac{1}{4}) f(\theta, \phi) \right],
\end{equation}
where $ k = \gamma/\sigma $, and we have defined $ f( \theta, \phi ) =
\csc2(\theta - \phi) (\sin \phi + \mu \sin 2 \phi) $.
We numerically minimize the free energy (\ref{Full-Free-Energy-vs-Phi})
with respect to $ \phi $ to find
the dependence of $ R $ on $ \mu_{24} $ (Fig.(\ref{R-vs-mu24})),
and of $ \phi $ on $ \mu_{24} $ (Fig.(\ref{Phi-vs-mu24})).

\begin{figure}
\includegraphics[width=6cm]{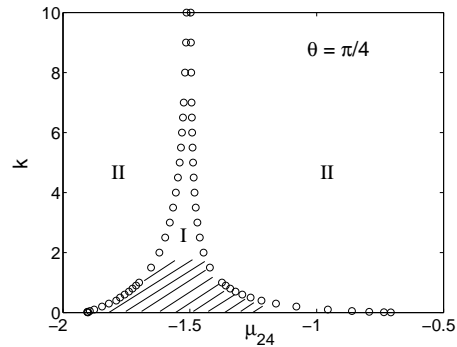}
\caption{\label{Phase-Diagram-k-vs-mu24} ``Phase diagram" in the $
k-\mu_{24} $- plane for $ \theta = \pi/4 $. Escape cylinders with a
finite radius are stable in the region labelled  I, and unstable in regions
labelled II. The range of stability is larger for smaller
values of $ k $, {\it i.e.} for stronger anchoring and weaker
nematic-isotropic surface tension. The continuum description holds
in the hashed region (see the discussion following
Eq.(\ref{Full-Free-Energy-vs-Phi})).}
\end{figure}

\begin{figure}
\includegraphics[width=6cm]{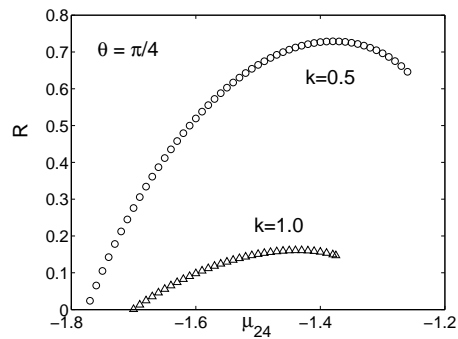}
\caption{\label{R-vs-mu24} Radius of the cylinder $ R $ measured
in units of $ \lambda_{24} = K_{24}/\sigma $ (which is about 0.3 $
\mu $m for a typical nematic) as a function of $\mu_{24}$ for $
\theta = \pi/4 $; $ k = 0.5 $ (circles) and $ k = 1 $ (triangles).}
\end{figure}

We now estimate the approximate value of the radius at which
double-tilt cylinders are energetically favored over the uniform,
undistorted nematic at the $ NI $ transition. With $ \mu_{24}
\approx - 1.5 $ and $ \lambda_{24} \approx 3 \times 10^{-5}$ cm;
the radius of the double-tilt cylinders is about $ 0.04 \, \mu$m
for $ k \approx 1 $, whereas it is about $ 0.1 \, \mu$m for $ k
\approx 0.5 $. It is important to note that the continuum
description that we have employed fails at very small radii of the
cylinders. The $ NI $ transition is first order in nature, and the
order parameter coherence length $\xi$ is of the order of
molecular dimensions (see \cite{de-Gennes:93}). The cylinder sizes
estimated above (for a reasonable range of parameter values) are
about ten times the order parameter coherence length. From
Fig.(\ref{R-vs-mu24}) we expect the continuum description to be
valid within the shaded region indicated in
Fig.(\ref{Phase-Diagram-k-vs-mu24}).

Fig.(\ref{Phase-Diagram-k-vs-mu24}) depicts the range of stability
(labelled I) of the double-tilt cylinders in the $ k - \mu_{24} $
plane. In the region labelled II on the right- hand side of the
spike in Fig.(\ref{Phase-Diagram-k-vs-mu24}) the gain in energy
from the saddle-splay term is not enough to overcome the energy
costs from the Frank free energy, the surface tension and the
anchoring energy terms. On the left hand side of the stable spike
in figure \ref{Phase-Diagram-k-vs-mu24}, as the {\it signed} value
of $K_{24}$ decreases, and $\phi$ at the boundary approaches
$\theta$ in order to gain energy from the saddle-splay term (see
Fig.(\ref{Phi-vs-mu24})). However, the torque balance condition
(\ref{Boundary-Condition}) at the boundary forces the relation
(\ref{Radius}), thereby reducing the radius of the cylinders (see
Fig.(\ref{Total-Free-energy})). Equation (\ref{Radius}) implies
that the radius of the cylinders goes to zero for $ 1 + \mu \cos
\phi \rightarrow 1 $. Finally we discuss the possibility of
forming lattices of escape cylinders. For a two dimensional
lattice, continuity of the director field across the planes
connecting the axes of neighboring double-tilt cylinders enforces
the condition that these have opposite escape directions. Clearly,
the triangular lattice is frustrated. The proposed square lattice
is antiferroelectric.
\begin{figure}
\includegraphics[width=6cm]{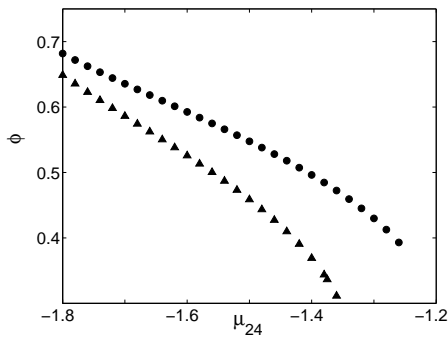}
\caption{\label{Phi-vs-mu24} The variation of  $ \phi $ with $
\mu_{24} $ for $ k = 0.5 $ (circles), $ k = 1.0 $ (triangles), for
$ \theta = \pi/4 $. As expected, $ \phi $ is closer to $ \theta $
for smaller values of $ k $ and $ \mu_{24} $.}
\end{figure}
Having demonstrated the feasibility of forming a two-dimensional
lattice of double-tilt cylinders for reasonable parameter values,
we note the following: ({\it i}) it may be possible to construct
three dimensional lattices out of escape configurations which are
not necessarily cylindrical in shape. We have explored the
possibility of forming three dimensional structures of right
circular double-tilt cylinders. Cubic lattices (BCC and FCC) can be 
formed if the angle $\phi = \pi/4$. For all other angles we
find that matching the director field at the boundaries of
neighboring cylinders introduces additional distortions in the
director field of the double-tilt cylinders. Energetics of these
three dimensional structures is beyond the scope of this paper.
({\it ii}) It is possible to construct lattices composed of {\em
hyperbolic} escape configurations (which correspond to escaped -1
disclinations), and ({\it iii}) the double-tilt cylinders can form
a molten line liquid analogous to the molten flux line
liquids\cite{Nelson:88} and the chiral line
liquids\cite{Kamien:93,Navailles:98} of liquid crystals. In
conclusion we have clearly demonstrated that the structure
proposed in this paper has a lower free energy than the nematic as
well as the isotropic phase near the $ NI $ transition. Whereas we
do not claim that this structure is the global minimum of the free
energy, our demonstration strongly suggests the possibility of
lattices composed of double-twist cylinders.

Freeze fracture electron microscopy of pure compounds exhibiting a
broad $NI$ transition may be a possible way to search for the
proposed phase, since the length scale involved ($0.05 \mu m$ to
$0.1 \mu m$) is inaccessible to other experimental probes. We urge
experimentalists to look closely for signatures of such structures
near the $NI$ transition.

We thank T. C. Lubensky for discussions, J. V. Selinger, G. S.
Ranganath, A. J. Levine and J. P. Sethna for useful suggestions.
BC thanks A. J. Levine for financial support. YH thanks NSF grant
no. DMR-0209256 for partial support.


\begin{thebibliography}{}
\bibitem[a]{a} Now at Department of Physics, Harvard University,
Cambridge, MA 02138.
\bibitem{de-Gennes:93} P. G. de-Gennes, and J. Prost {\it The physics of Liquid
Crystals}, Clarendon Press, Oxford, 1993; and references therein.
\bibitem{Meiboom:81} S. Meiboom, J. P. Sethna, P. W. Anderson and
W. F. Brinkman, Phys. Rev. Lett. {\bf 46}, 1216 (1981).
\bibitem{Renn:88} S. R. Renn and T. C. Lubensky, Phys. Rev. A {\bf
38}, 2132 (1988).
\bibitem{Goodby:88} J. Goodby, M. A. Waugh, S. M. Stein, E. Chin, R. Pindak,
and J. S. Patel, Nature {\bf 337}, 449 (1988); J. Am. Chem. Soc. {\bf
111}, 8119 (1989).
\bibitem{DiDonna:03} B. A. DiDonna and R. Kamien, Phys. Rev. E
{\bf 68}, 041703 (2003).
\bibitem{Pansu:01} E. Grelet, B. Pansu, M-H. Li, and H. T. Nguyen, Phys. Rev. Lett. {\bf 86}, 3791
(2001).
\bibitem{Mayer:73} R. B. Mayer, Phil. Mag. {\bf 27}, 405 (1973).
\bibitem{Cladis:72} P. Cladis and M. Kleman, J. Phys. (Paris) {\bf 33}, 591 (1972).
\bibitem{footnote2} The ``one-constant" approximation is valid in as much as
the contribution proportional to the square of the nematic order parameter
does not contribute significantly to $ K_{1}, K_{2}$ and $ K_{3} $ near the
$ NI $ transition. See \cite{de-Gennes:93} above.
\bibitem{Rapini:69} A. Rapini and M. Papoular, J. Phys. (Paris) Colloq. {\bf 30}, C4-54 (1969).
\bibitem{Faetti-Palleschi:84} S. Faetti and V. Palleschi, Phys Rev. A {\bf 30}, 3241 (1984);
J. Phys. (Paris) Lett. {\bf 45}, L313 (1984).
\bibitem{Barbero:86} G. Barbero, I. Dozov, J. F. Palierne, and G. Durand, Phys. Rev. Lett.
{\bf 56}, 2056 (1986).
\bibitem{Prost:77} J. Prost and J. P. Marcerou, J. Phys (Paris) {\bf 38}, 315 (1977).
\bibitem{Doane:92} G. P. Crawford, D. W. Allender and W. Doane,
Phys. Rev. A {\bf 45}, 8693 (1992).
\bibitem{Nelson:88} D. R. Nelson, Phys. Rev. Lett. {\bf 60}, 1973
(1988); D. R. Nelson and H. S. Seung, Phys. Rev. B, {\bf 39}, 9153
(1989).
\bibitem{Kamien:93} R. D. Kamien and T. C. Lubensky, J. Phys. I France {\bf 3}, 2131
(1993).
\bibitem{Navailles:98} L. Navailles, B. Pansu, L. Gorre-Tallini,
and H. T. Nguyen, Phys. Rev. Lett. {\bf 81}, 4168 (1998).
\end{thebibliography}
\end{document}